# Results of High Order Modes Spectra Measurements in 1.3 GHz Cavities for LCLS-II

A. Lunin[†], T. Khabiboulline, A. Sukhanov and V. Yakovlev, Fermilab, Batavia, USA

*Abstract*

Fermilab recently completed production and testing of 1.3 GHz cryomodules for the LCLS-II project. Each cryomodule consists of eight TESLA-shaped superconducting elliptical cavities equipped with two High Order Mode (HOM) coupler ports. Measurement of the HOM spectrum is part of the incoming quality control of cavities at room temperature and the final qualification cold test of cryomodules at the Cryomodule Test Facility (CMTF). In this paper we describe the procedure for measuring the HOM spectrum along with further data processing. Finally, we present accumulated statistics of individual HOM frequencies and quality factors related to various cavity vendors and discuss the possible contribution of HOMs to heat loads and beam dynamics.

## Introduction

Advances in superconducting technologies result in realization of many projects of particle accelerators operating both in continuous mode and in modes with high duty factors of the beam current (XFEL, LCLS-II, ILC). Superconducting (SRF) cavities are a good resonant system with a spectrum of modes having very high-quality factors. A beam of charged particles can potentially cause resonant excitation of a particular High Order Mode (HOM) and lead to both large beam power losses and dilution of the beam emittance. Due to the nature of SRF cavities, they have a random eigen spectrum, and coherent HOM excitation becomes an inherently probabilistic problem that is very difficult to evaluate [1]. Thus, only direct measurement of the HOM parameters under operating conditions makes it possible to obtain a reliable model of the resonant HOMs excitation a beam of an arbitrary pattern. In addition, monitoring HOM spectra allows any deviations to be identified earlier and notifies the supplier of non-conforming cavity fabrication.

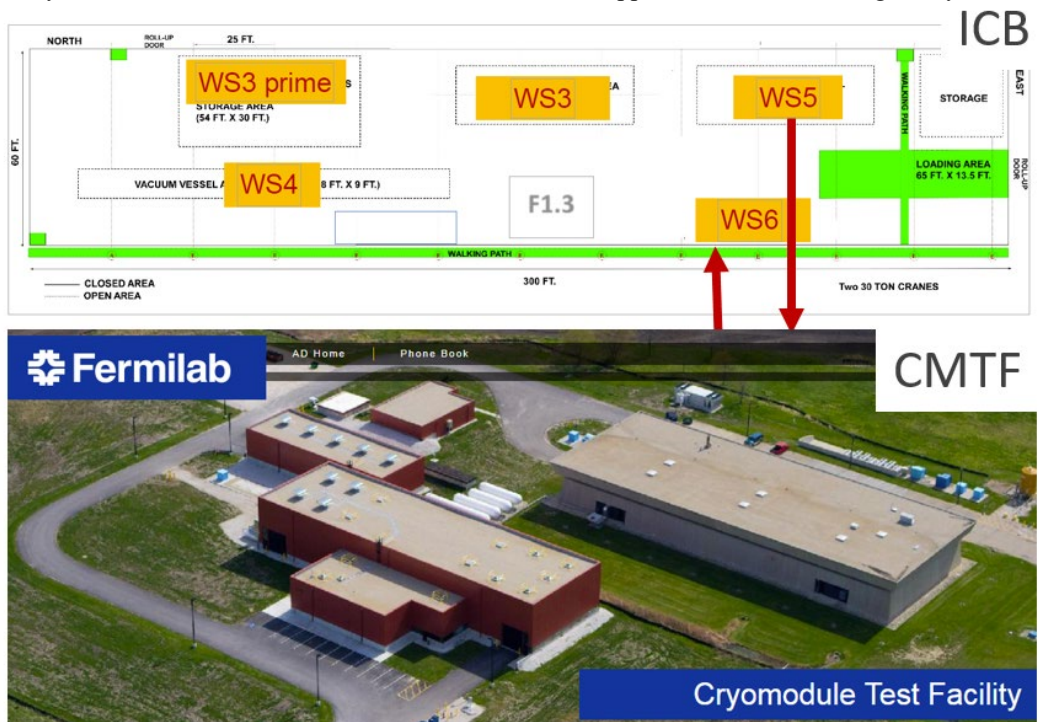

Fig. 1 LCLS-II cryomodule assembly workflow on the sequence of workstations (WS) and final RF testing at Fermilab

As part of the LCLS-II collaboration, Fermilab was responsible for the production and testing of twenty 1.3 GHz cryomodules [2]. For technological purposes, two independent companies (vendor A and vendor B) with approximately equal contributions were selected to manufacture SRF cavities. When a cavity is received from a vendor, it follows standard cryomodule assembly procedures, including a series of incoming quality checks (QC). Figure 1 shows the workflow for final assembly of the LCLS-II cryomodule in the high-bay ICB building and subsequent high power RF testing at the dedicated Cryomodule Test Facility (CMTF) at Fermilab. Measurements of the HOM spectra of the cavities were taken during tests in the CMTF after the cooling the cryomodule to 2K and tuning all fundamental power couplers (FPC) to a nominal external coupling of 4E7. After post-processing the raw data, the sorted HOM parameters are recorded in the online traveler database for each delivered cryomodule.

* This work was produced by Fermi Research Alliance, LLC under Contract No. DE-AC02-07CH11359 with the U.S. Department of Energy. Publisher acknowledges the U.S. Government license to provide public access under the DOE Public Access Plan
† lunin@fnal.gov

## RF Measurements of HOM Spectra

The layout of the 9-cell 1.3 GHz cavity for the LCLS-II project is illustrated in Fig. 2. The cavity has an FPC port and three auxiliary ports, including two downstream and upstream HOM couplers. To measure the HOM spectrum, we recorded signal transmission from HOM1 ports to HOM2 ports. Since the cryomodule is being tested inside a concrete cave used as radiation shielding, there are approximately 20 meters of ⅜-inch diameter Heliax coaxial cables connecting the auxiliary ports to the low-power electronics rack located outside the cave. These HOM cables were temporarily used to connect to the vector network analyzer (VNA) ports to measure the S21 transmitted signal. Typical insertion loss in cables is about 3 dB, which is practically negligible compared to the dynamic range of the VNA of more than 120 dB [3].

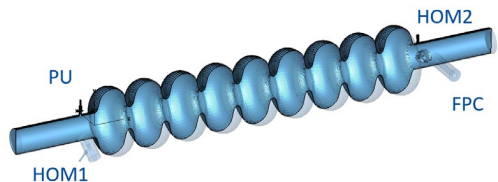

Fig. 2 1.3 GHz LCLS-II cavity with four axillary ports: field pickup (PU), two HOM couplers and FPC

An example of the recorded cavity spectrum up to 4 GHz is shown in Fig. 3 together with the normalized shunt impedances of the most dangerous HOMs. One can see that the spectrum above 2.5 GHz becomes populated with overlapping cavity passbands. In the latter case, there is no practical method for identifying a single mode, since the stochastic frequency deviation is much larger than the distance between the modes. Therefore, to collect statistical data, we measured only the lowest passbands containing HOMs with the highest R/Q values. There are a total of 45 modes in two dipole passbands (1.6 GHz to 1.9 GHz) and a second monopole passband (2.35 GHz to 2.45 GHz). In addition, for a fraction of the cavities we also recorded spectra of the first monopole (from 1.27 to 1.3 GHz) and quadruple (from 2.29 to 2.32 GHz) passbands. For automatic data acquisition, LabVIEW code was developed that performs frequency scans in several sub bands with different local frequency steps depending on the anticipated HOM quality factor. This allows us to significantly reduce the amount of data and the acquisition time.

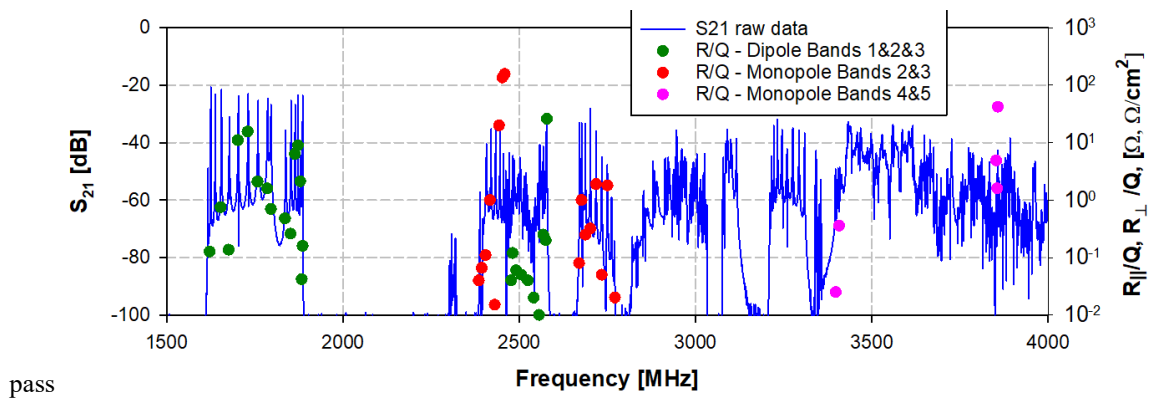

pass

Fig. 3 Measured cavity S21 signal (HOM1 to HOM2) and the shunt impedances for different HOM passbands

Data processing is performed using programming in the Mathcad environment, which automatically finds peaks above a given threshold and calculates quality factors [4]. In the case of dipole modes, there is a possibility that the frequencies of the two polarizations of the same mode will be very close to each other, and the two peaks will overlap in the resulting transmission signal. The remedy is a multidimensional fit where the parameters of the two modes (frequency, amplitude and relative phase shift) are scanned to match the measured S21 signal. The result of the fitting algorithm is shown in Fig. 4, where the blue and green curves are the reconstructed resonances of two different polarizations.

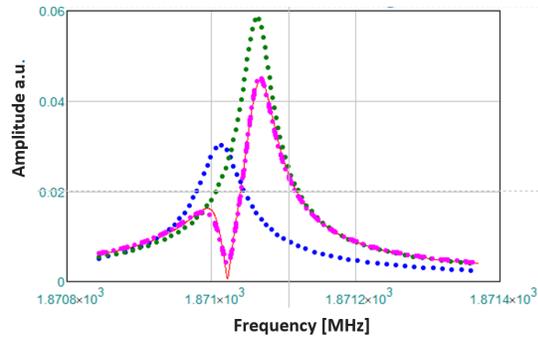

Fig. 4 Decomposition of two nearby dipole modes (solid red is raw data, dashed pink is fitted curve and dotted blue and green are restored resonance peaks)

## Statistical Results of HOM Spectra Measurements

After post-processing is completed, the HOMs are sorted, and the HOM parameters are entered into the CM traveler. Typical HOM spectra of eight cavities installed in the cryomodule are shown in Fig. 5 for the first dipole and monopole passbands.

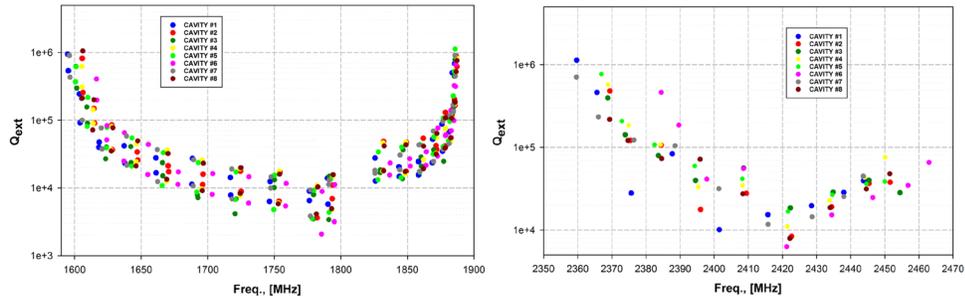

Fig. 5 HOM spectra of 8 cavities in the cryomodule, dipole (left) and monopole (right) passbands.

The production of 1.3 GHz cryomodules for the LCLS-II project was completed in 2022, with the delivery of the last cryomodule to SLAC. In total, we collected data for 320 cavities manufactured by vendors A and B, respectively. Statistical analysis of the HOM data revealed significant deviations in frequencies and quality factors compared to parameters calculated for ideal cavity geometry. A comparison of the standard deviations of the first monopole passband frequencies and the errors of the cavity lengths is shown in Fig. 6 for vendors A and B respectively. This result is consistent with the theoretical estimate made for similar cavities in the XFEL project [5]. Note that the systematic difference between the average frequencies of the two vendors is much larger than the deviations themselves, which is explained by the variations in the cavity length of ±2 mm.

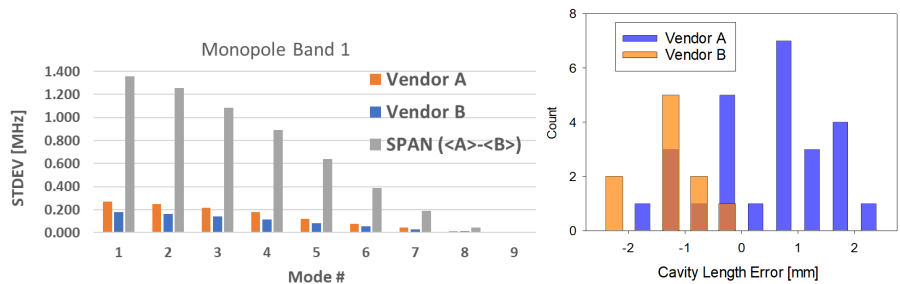

Fig. 6 Standard frequency deviations of the first monopole passband (left) and the length errors (right) of cavities produced by vendors A and B respectively.

Finally, we calculated statistics for all HOMs in the five lower passbands, which is presented in Fig. 7. On average, the standard deviations of HOM frequencies increase approximately linearly with the offset from the operating frequency, while within a given passband they can vary several times, since individual HOMs have different sensitivities to changes in apertures or lengths of the cavity cells. The most important are the deviations of the HOM quality factors, since they can significantly affect the beam dynamics. For some modes, the highest measured Q-loaded values are several times larger than those in simulations, mainly because in the worst scenario, deformations of the end cells can reduce the coupling of the inner cells to the beam pipe and thus significantly increase the external quality factor. As a result, in rare cases the Q-loaded can be ten times larger than expected.

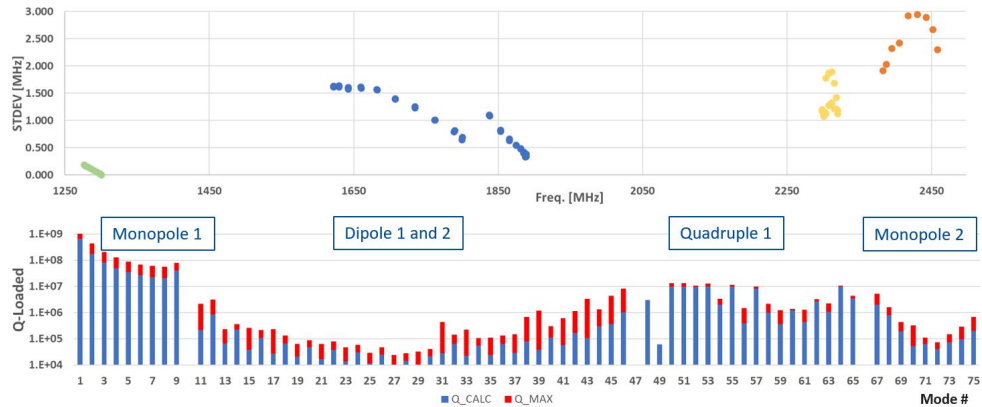

Fig. 7 Measured HOM statistics for five passbands of the 1.3 GHz LCLS-II cavity, standard frequency deviations (upper plot) and maximum (red) and simulated (blue) quality factors (bottom plot)

## Conclusions

In a frame of the LCLS-II collaboration, Fermilab produced and delivered twenty 1.3 GHz cryomodules to SLAC. During the production process, we closely monitored the RF parameters of the SRF cavities, including the HOM spectra of the first five cavity passbands up to 2.5 GHz. The measured HOM statistics revealed significant variations in frequencies and quality factors compared to the simulation. On average, the HOM frequency deviations scale linearly (about 0.3% per GHz) with respect to the offset from the operating mode, while it also has significant variations within the passband. The measured HOM quality factors can be ten times larger than their nominal values.

In overall, the detailed statistics of HOM parameters presented allows modeling the resonant excitation of HOM by an arbitrary beam pattern, which can significantly improve the reliability of future accelerator designs, especially with high average beam current.